\documentclass[twocolumn]{jpsj2}
\usepackage{epsfig}
\def\gsim{\; $\raise0.3ex\hbox{$>$}\llap{\lower0.8ex\hbox{$\sim$}}$\;}
\def\lsim{\; $\raise0.3ex\hbox{$<$}\llap{\lower0.8ex\hbox{$\sim$}}$\;}

\title{Magnetic phase diagram of the S=1/2 antiferromagnetic zigzag spin chain in the strongly frustrated region: cusp and plateau}
\author{Kouichi Okunishi and  Takashi Tonegawa${}^1$}
\inst{
Department of Physics, Faculty of Science, Niigata University, Igarashi 2, 950-2181, Japan\\
${}^1$Department of Mechanical Engineering, Fukui University of Technology, 6-1 Gakuen 3-Chome, Fukui-Shi 910-8505, Japan.
}
\date{\today}

\abst{ We determine the magnetic phase diagram of the
$S=1/2$ antiferromagnetic zigzag spin chain in the strongly frustrated
region, using the density matrix renormalization group method.  We
find the magnetization plateau at 1/3 of the full moment accompanying
the spontaneous symmetry breaking of the translation, the cusp
singularities above and/or below the plateau, and the even-odd effect
in the magnetization curve. We also discuss the formation mechanisms
of the plateau and cusps briefly.}

\kword{frustration, cusp, 1/3 plateau, even-odd effect}

\begin{document}
\maketitle

For the purpose of clarifying the role of the frustration in low-dimensional
quantum spin systems, the $S=1/2$ antiferromagnetic zigzag
spin chain has been attracting considerable attention, since it minimally
contains the frustrating interaction without loss of the
translational invariance\cite{halzig,tone1,nomura,whiteaffleck}.  The
Hamiltonian of the model is given by
\begin{eqnarray}
{\cal H}&=& \sum_{i}[J_1 \vec{S}_{i}\cdot\vec{S}_{i+1} +J_2
\vec{S}_{i}\cdot\vec{S}_{i+2} ] - H \sum_i S_i^z,
\label{zigzag}
\end{eqnarray}
where $\vec{S}$ is the $S=1/2$ spin operator, $H$ is the magnetic
field, and $J_1$ and $J_2$ denote the nearest and next nearest
neighbor couplings, respectively. We introduce the notation
$\alpha=J_2/J_1$ for simplicity.  Recently, the zigzag chain was
realized as SrCuO$_2$\cite{matsuda}, Cu(ampy)Br$_2$\cite{kikuchi},
(N$_2$H$_5$)CuCl$_3$\cite{haginaru} and F$_2$PIMNH\cite{hosokoshi}.

The Hamiltonian (\ref{zigzag}) has a very simple form, but captures a
variety of behaviors induced by the frustration.  In particular, the
zigzag chain in a magnetic field has been studied
actively\cite{tone2,schmidt,cabra,cusp,maeshima} and it has been clarified
that cusp singularities appear near the saturation field and/or in the
low field region for $\alpha \le 0.6$, in accordance with the
frustration-driven shape change of the dispersion curve of the
elementary excitations\cite{cusp,maeshima,mobius}.  However, the
magnetic phase diagram including the strongly frustrated
region ($\alpha >0.6$) has not been acquired yet.  Here we remark that
the magnetization curve of $\alpha=0.6$ is still quite different from
the one in the $\alpha\to \infty$ limit (see Fig.\ref{fmh1} (a)).  Thus, as
$\alpha$ is increased beyond $\alpha=0.6$, the intrinsic structural
change of the magnetization curve can be expected particularly around
$\alpha \simeq 1$, where the most significant competition is achieved.

\begin{figure}[ht]
\begin{center}
\epsfig{file=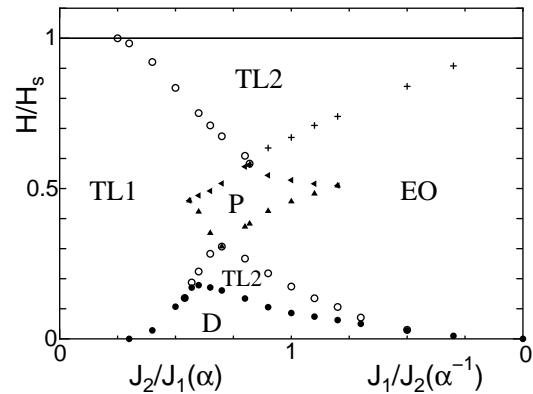,width=7cm}
\end{center}
\caption{Magnetic phase diagram of the zigzag spin chain.  D: dimer
gapped phase, P:1/3 plateau, TL1: one component TL liquid, TL2: two
component TL liquid, and EO: even-odd behavior branch.  The open
circles denote the position of the cusp singularities. The solid
circles represent the boundary of the dimer gapped phase. The triangles represent the upper and lower edges of the 1/3 plateau, and the crosses indicate
the upper edge of the even-odd effect in the magnetization curve. The
saturation field is normalized to be unity.} \label{phase}
\end{figure}

\begin{figure}[htb]
\begin{center}
\epsfig{file=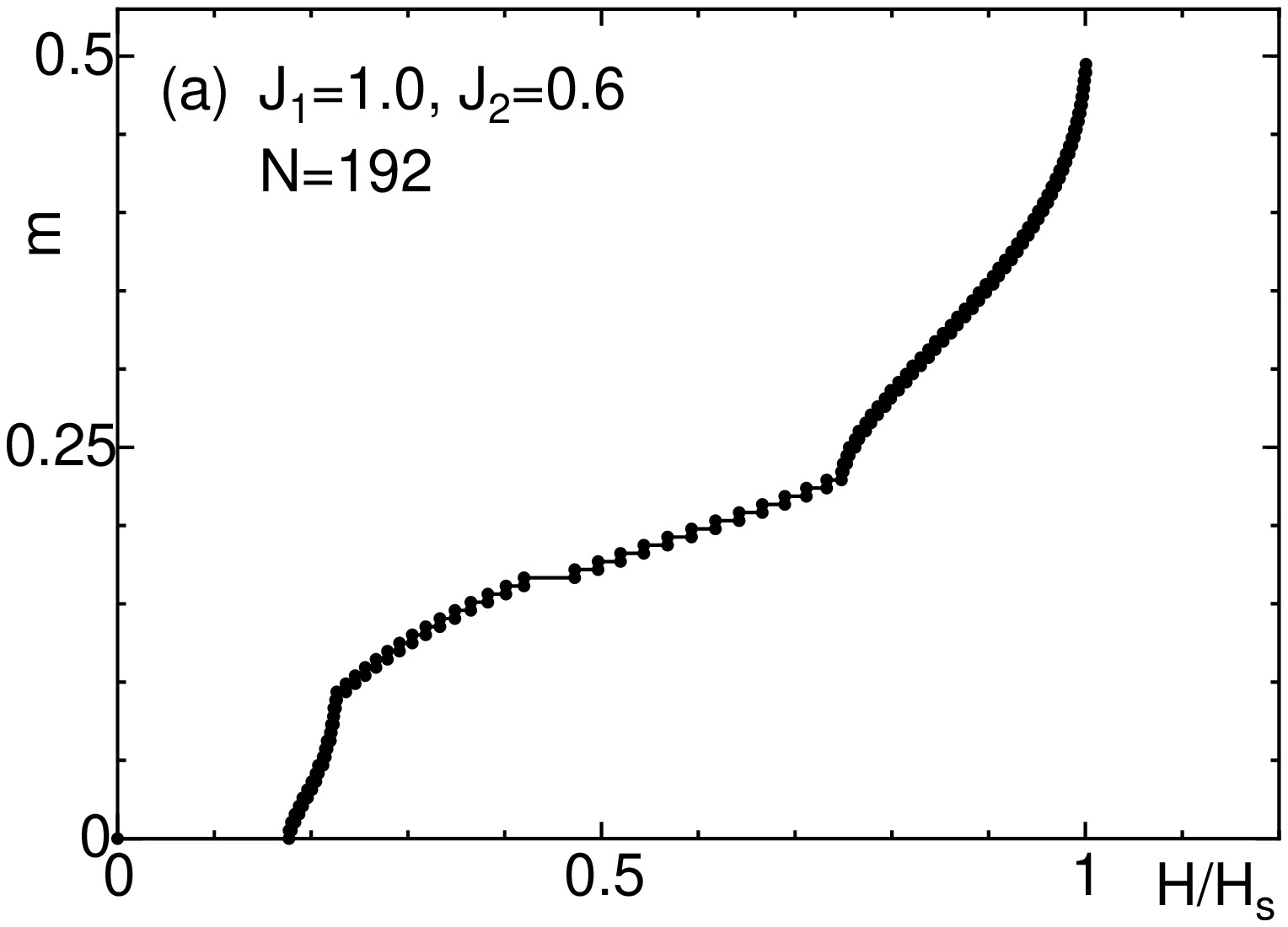,width=7cm} \epsfig{file=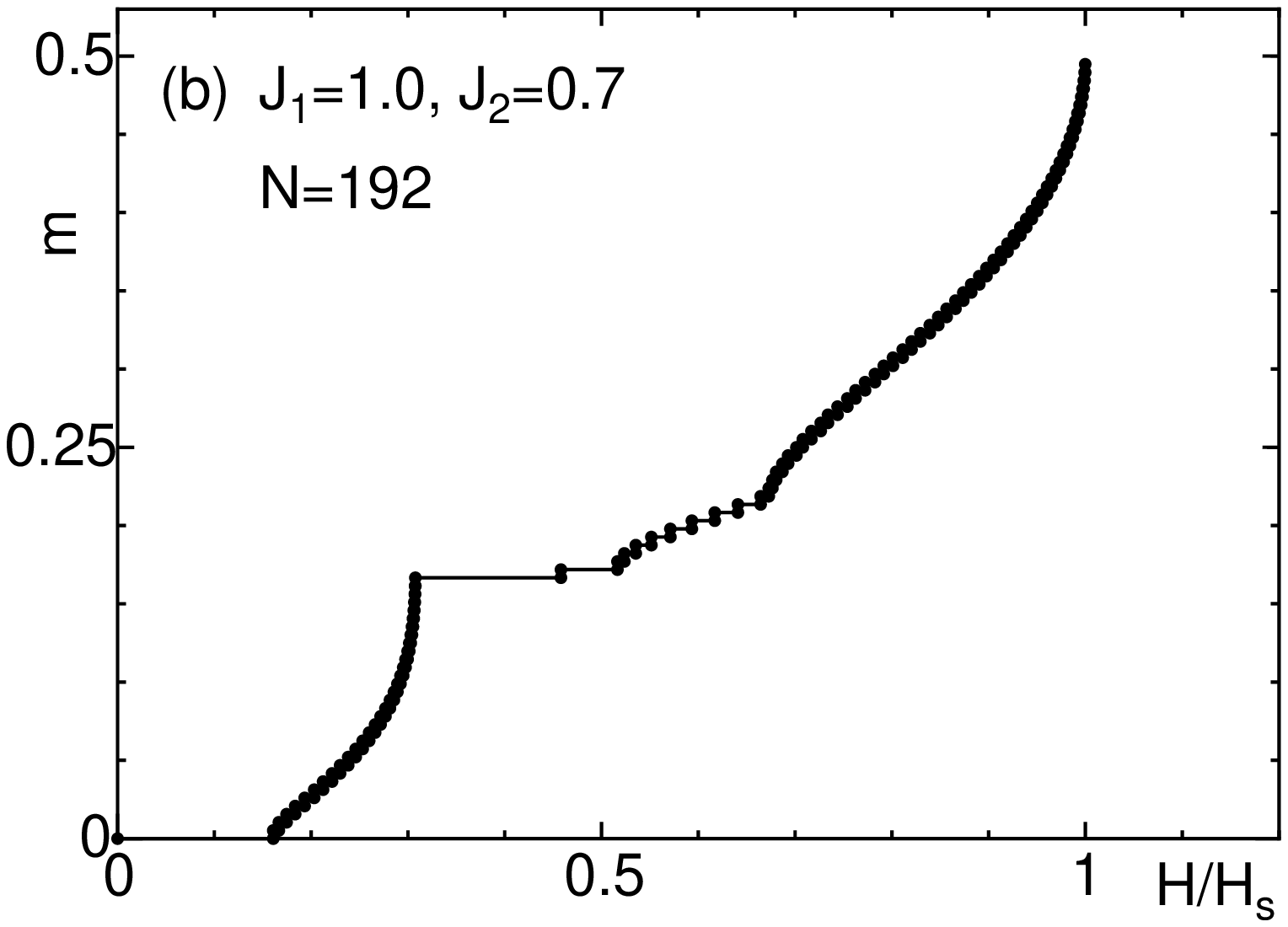,width=7cm}
\epsfig{file=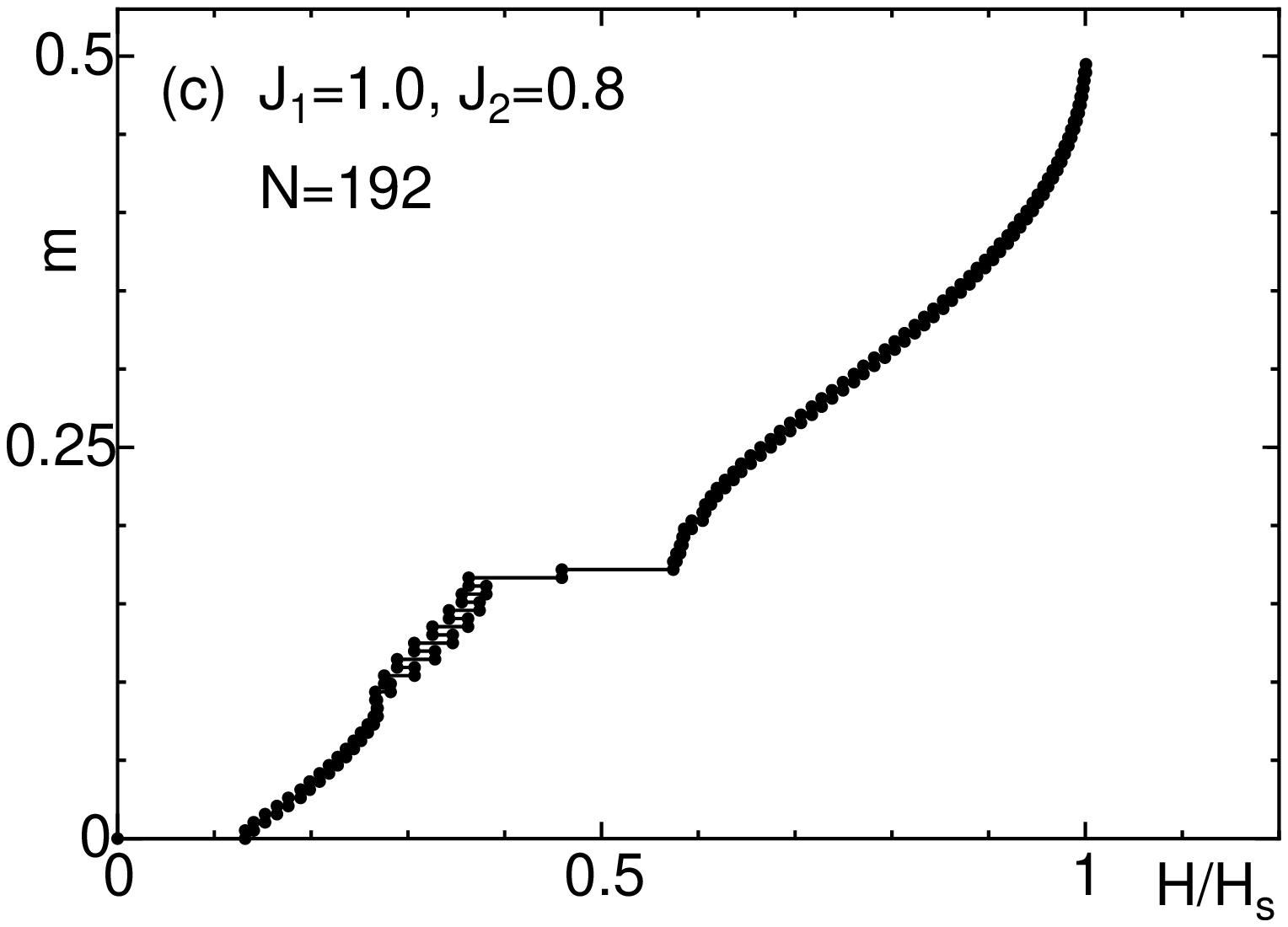,width=7cm}
\caption{Magnetization curves for $\alpha=0.6$(a), 0.7(b) and 0.8(c).
The vertical axis indicates $m\equiv M/N$.}\label{fmh1}
\end{center}
\end{figure}

In this paper we address the magnetization process of the zigzag chain
in the strongly frustrated region($\alpha>0.6$), using the density
matrix renormalization group (DMRG) method\cite{dmrg}.  We then find
that the obtained phase diagram exhibits rich physics, as is
shown in Fig.\ref{phase}.  For $0.56\lsim \alpha\lsim 1.25$, the
magnetization plateau appears at 1/3 of the full moment, accompanying
the spontaneous breaking of the translational symmetry with the  period
three.  Moreover, the cusp singularities in the magnetization curve
show quite interesting behavior; as $\alpha$ is increased, the
high field cusp merges into the 1/3 plateau at $\alpha \simeq
0.82$. Also the low field cusp merges the 1/3 plateau at $\alpha
\simeq 0.7$, but it appears again when $\alpha > 0.7 $.  In addition,
we find an interesting even-odd effect in the magnetization curve for
$\alpha > 0.7 $.

In analyzing the phase diagram, an important feature of the zigzag
chain is that the Hamiltonian (\ref{zigzag}) interpolates between the
single Heisenberg chain($J_2=0$) and the double Heisenberg
one($J_1=0$) continuously.  In other words, we can describe the
frustration effect in the zigzag chain as the interplay between the
single and double chain natures.  Although such a view point has not
been emphasized so far, it provides some essential insights on the
frustration effect in the magnetization curve.  In fact, the cusp
transition for $\alpha<0.7$ mentioned above is described as the
transition between the one and two component Tomonaga-Luttinger(TL)
liquid\cite{cusp}; the two component TL liquid is realized below the
lower cusp and/or above the upper cusp reflecting the two chain nature
of the system, while the middle-field branch still consists of the one
component TL liquid.  In the following, we show the results of the
magnetization curves for $\alpha>0.6$ and then proceed to a detailed
analysis of each phase.  How can we connect the above noted
competition with the characteristic behaviors in each phase?

{\it magnetization curves}: In order to compute the magnetization
curve, we employ the DMRG of both the finite system size and infinite
system size versions.  However, here we mainly present the results for 
the finite size systems.  We compute the minimum energy $E_N(M)$ for a
magnetization $M$ of the $N=192$ site system with the free boundary
condition, and then obtain the magnetization curve by determining the
level crossing point from $E_N(M)$ and $E_N(M\pm 1)$.  The maximum
number of the retained bases in the DMRG computation is typically
$60$, where the energy is converged sufficiently.

In Fig.\ref{fmh1} (a) we show the magnetization curve for
$\alpha=0.6$, where the scale of the magnetic field is normalized with
the saturation field $H_s$.  The cusp singularities at $H/H_s \simeq
0.22$ and $ H/H_s \simeq 0.75$ are the same as those reported in the
previous papers, which are explained well by the shape change of the
spinon or spin wave dispersion curves.\cite{maeshima,mobius} A novel
point in the figure is that the 1/3 plateau is found to appear
accompanying the spontaneous symmetry breaking of the translational
invariance.  The lower critical field of this plateau is
$H_{c1}/H_s=0.424$ and the upper critical field is $H_{c2}/H_s=0.477$.
The details of the 1/3 plateau are discussed later.

As $\alpha$ is increased, the cusps approach the plateau, which
implies that the two component TL region becomes more dominant, and at
the same time the width of the plateau extends.  Figure \ref{fmh1} (b)
shows the magnetization curve of $\alpha=0.7$, in which the width of
the 1/3 plateau is extended to 0.209.  An anomalous step corresponding
to $M=33$ in the 1/3 plateau is due to the open boundary effect.  By
analyzing the size dependence, we have confirmed that this step is
almost independent of $N$ and is thus negligible in the bulk limit.  In
addition to the plateau, an interesting point is that the low field
cusp is merged into the 1/3 plateau, and then the branch below the
plateau sticks into the plateau with infinite slope. On the other
hand, the high field cusp still survives.

In Fig.\ref{fmh1} (c), we show the result for $\alpha=0.8$.  In the
figure we can clearly see that the high field cusp almost merges
into the $1/3$ plateau; a precise computation indicates that the
disappearing point of the high field cusp is $\alpha\simeq 0.82$.  On
the other hand, a cusp singularity appears again in the low field
region and, moreover, the magnetization curve above this low field
cusp remarkably exhibits an anomalous even-odd oscillation with
respect to the magnetization $M$.  
Of course,  the anomalous steps corresponding to $M$=odd  must be skipped 
in the true magnetization curve of the finite size system which is determined  from $E_N(M)$ with $M=$even\cite{eo}.
However, we purposely show these anomalous odd steps,
since they are reflecting an interesting aspect of the double-chains
nature of the system.  We discuss the details of this even-odd effect
later, but here simply note that the magnetization curve obtained with the
infinite system DMRG is consistent with the present 192-site result.

\begin{figure}[ht]
\begin{center}
\epsfig{file=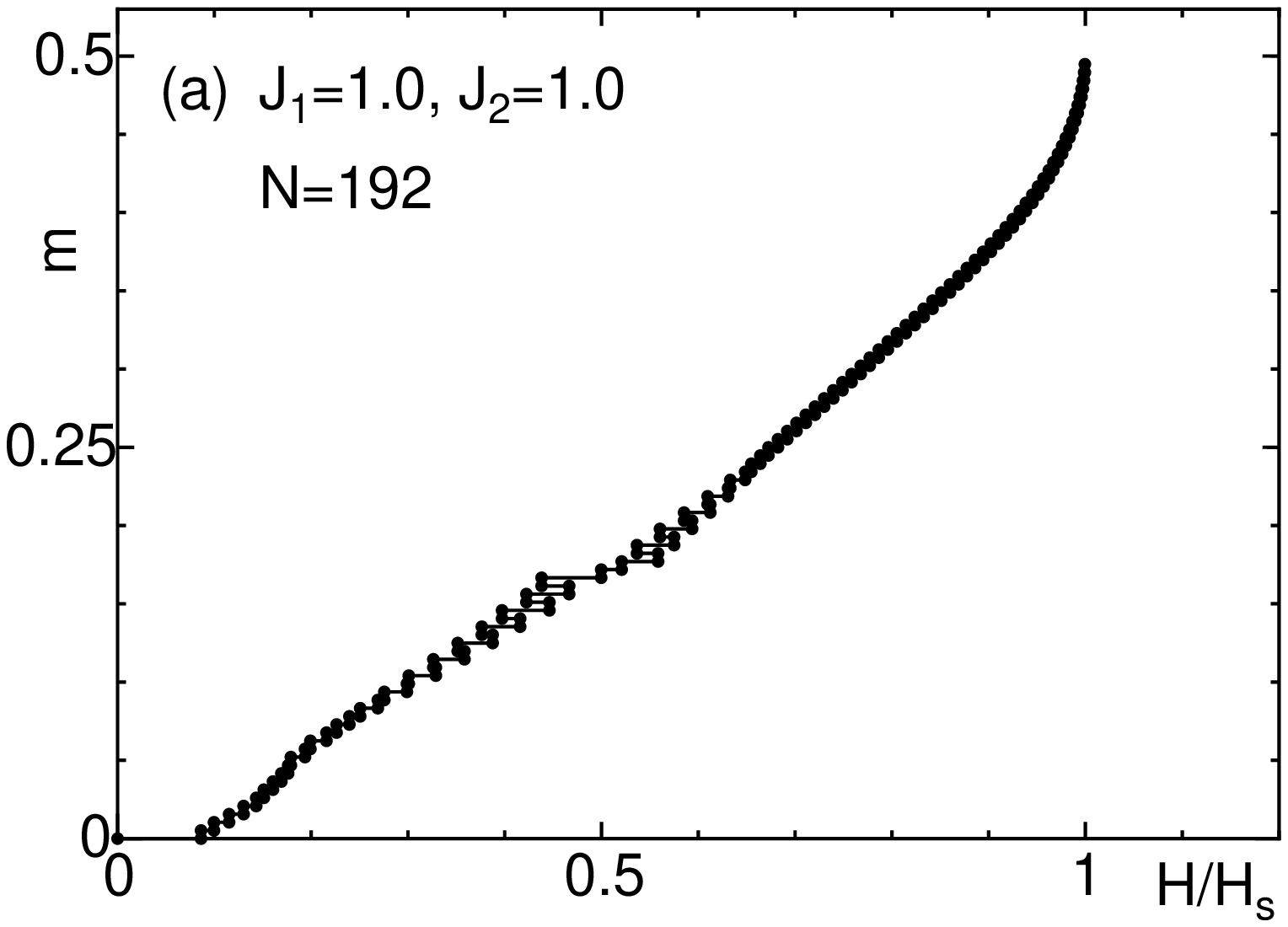,width=7cm}

\epsfig{file=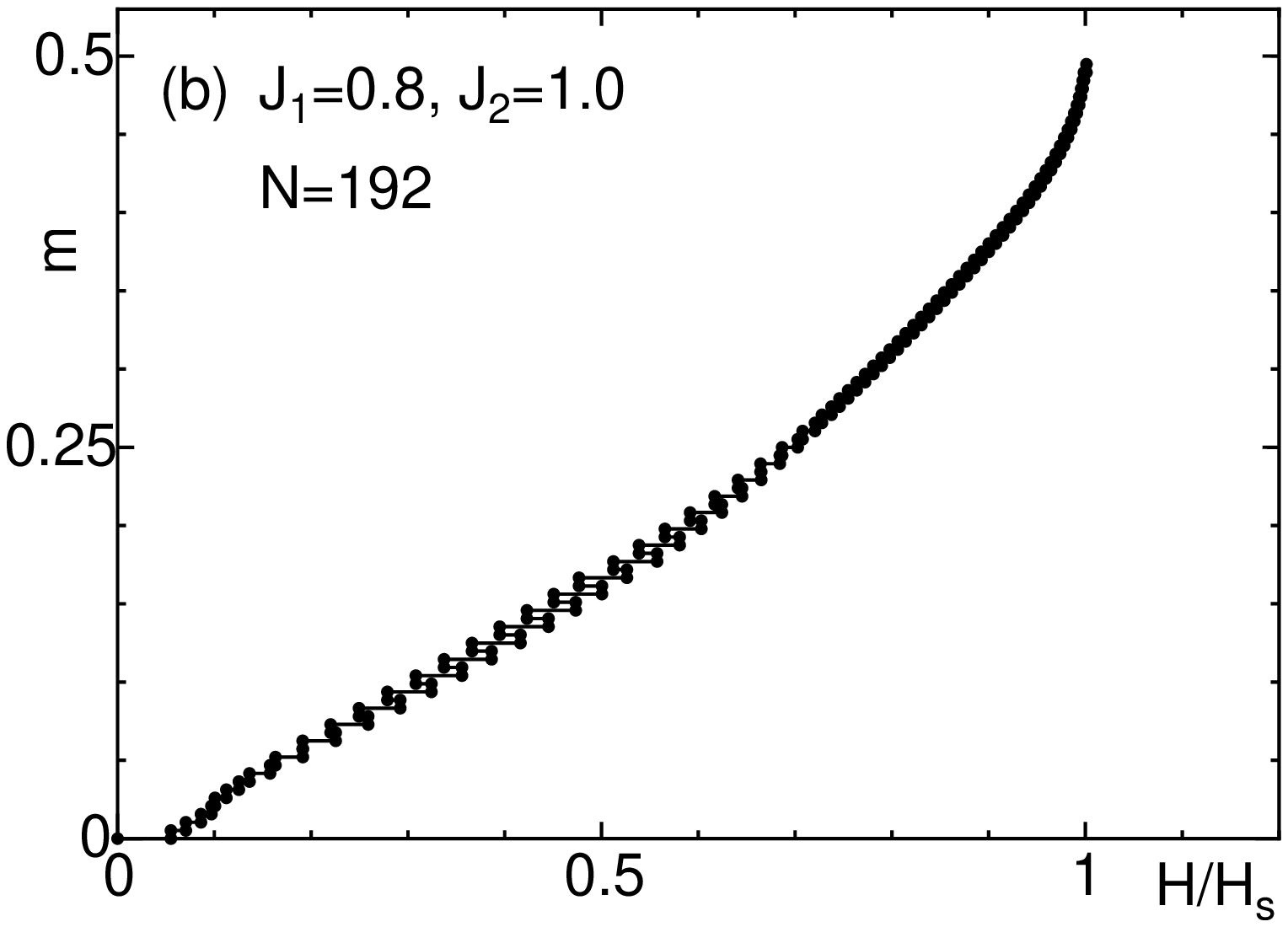,width=7cm}
\end{center}
\caption{Magnetization curves for $\alpha=1.0$(a) and
$\alpha^{-1}=0.8$(b).}\label{fmh2}
\end{figure}

As $\alpha$ is increased beyond $\alpha=0.82$, the 1/3 plateau turns
to shrink.  We show the results for $\alpha=1.0$ in Fig.\ref{fmh2}
(a), where the plateau becomes narrower compared with that for the $\alpha=0.8$
result and the cusp shifts to the lower field side.  However, the
high field cusp does not appear again.  Instead, the even-odd
behavior of the magnetization curve is extended above the 1/3 plateau.
This upper edge of the even-odd effect shifts to the higher field side
as $\alpha$ is increased.  Although there is a possibility of the cusp
appearing at the upper edge of the even-odd branch, we can not confirm
it within the present numerical results for 192 sites.

Figure \ref{fmh2}(b) shows the magnetization curve of
$\alpha^{-1}=0.8$.  In the figure, the region of the even-odd
oscillation is extended to a higher field region and the plateau
becomes indistinct.  Indeed, for $\alpha^{-1}=0.8$, the infinite system
DMRG illustrates a continuous curve above the low filed cusp up to
the saturation field, implying that there is no $1/3$ plateau.  As
$\alpha$ is increased further, the branch above the cusp becomes more
similar to that of the (double) Heisenberg chain.  At the same time,
the low field cusp itself approaches  the dimer gap boundary.  In the
$J_1\to 0(\alpha^{-1}\to 0)$ limit, the ``amplitude'' of the even-odd
oscillation becomes smaller and finally the magnetization curve becomes
equivalent to that of the Heisenberg chain again.  However, we note
that it is difficult to see how the cusp behaves in this limit simply by the
direct observation of the magnetization curve.

\begin{figure}[ht]
\begin{center}
\epsfig{file=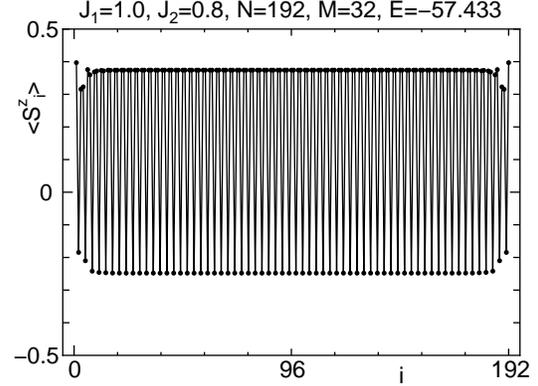,width=7cm}
\end{center}
\caption{ The distribution of the local spin moment at $M$=32(1/3
plateau) for $\alpha=0.8$}\label{spindistribution}
\end{figure}

{\it 1/3 plateau state}: Let us discuss the nature of the 1/3 plateau.
According to the quantization condition of the magnetization
plateau\cite{oshikawa}, the 1/3 plateau state satisfies
$q(1/2-M)=$integer with $M=1/6$, where $q$ is the period of the
plateau state.  This implies that the translational symmetry of the
Hamiltonian (\ref{zigzag}) must be spontaneously broken with
$q=3,6,\cdots$.  In order to see it, we calculate the local spin moments
at the plateau along the chain direction. The result is shown in
Fig.\ref{spindistribution}, where the value of the local moment is
developed to about $\langle S^z_i\rangle \simeq 0.38$ or $-0.26$ and
thus the ``up-up-down'' type long-range order is realized.  Moreover,
the boundary effect decays rapidly, which supports that the 1/3
plateau state has the excitation gap.

What is the origin of this up-up-down structure?  We have also
investigated the magnetization curve of the zigzag XXZ chain with the
Ising-like anisotropy, and then found that the 1/3 plateau is extended
up to the Ising limit.  Here it should be remarked that the 1/3
plateau of the up-up-down array is proven for the Ising anisotropic
limit\cite{morihori}, where the up-up-down structure is especially
compatible with the triangle structure of the zigzag chain.  In
addition, we have also checked that a similar 1/3 plateau of the
up-up-down array appears for the Ising-like classical XXZ zigzag chain
reflecting the triangular structure.\cite{cg} However, this
``classical'' plateau vanishes at the isotropic XXX chain.  Thus the
present 1/3 plateau is induced by the quantum effect.

\begin{figure}[ht]
\begin{center}
\epsfig{file=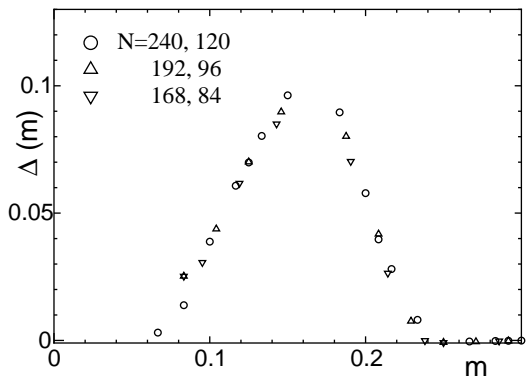,width=7cm}
\end{center}
\caption{ The binding energy for $\alpha=1.0$ obtained by the size extrapolation: $\Delta_N(m)\sim \Delta(m)+ {\rm const} /N $.  The open circles denote the result for $N=$240 and 120. The upward and downward triangles denote those for $N=$192 and 96, and  168 and 84 respectively. These results are in good agreements with each other.}\label{bind}
\end{figure}

{\it even-odd effect}: The anomalous even-odd behavior for
$\alpha>0.7$ can be explained in terms of the formation of the bound
state of magnetic quasi particles as follows. The spin chain with a
finite magnetization is described by the quasi particles carrying
$S^z=1$ distributed on the chain\cite{qp}.  When $\alpha$ is small and
the single chain nature is dominant, the magnetic particle is
distributed on the single chain, but if $\alpha$ becomes large and the
double chain nature becomes significant, the particles should be
distributed on each chain equivalently.  However, if the total
magnetization is odd, the remaining one particle is sitting on one of
the two chains.  Thus the $M$=odd system is less stable than the
$M$=even one, which suggests the formation of the bound state in the
thermodynamic limit.  
In order to see it, we define the binding energy
of the $N$-site system as $\Delta_N(m)\equiv 2( E_N(M+1)-E_N(M) )-
( E_N(M+2)- E_N(M) ) = 2E_N(M+1) - E_N(M)- E_N(M+2)$
for $M=$even.  Figure \ref{bind} shows the obtained result with the
size extrapolation of $\Delta_N(m)\sim \Delta(m)+ {\rm const} /N $
for $N=240, 192, 168, 120, 96, 84$,\cite{size} which shows clearly that the
binding energy develops in the region corresponding to the even-odd
behavior.  Here we note that this size extrapolation does not work
well around the low field cusp($m\simeq 0.6$) due to the singular
behavior.

The formation of the bound state of the quasi particles provides an
essential insight into the cusp mechanism for $\alpha>0.7$.  The
magnetization curve below the low field cusp consists of the 
two-component TL liquid that is continuously connected from the $\alpha
<0.7$ region, while the branch above the low field cusp consists of
the bound state of the magnetic particles.  Thus the cusp mechanism
for $\alpha>0.7$ is described by the transition from the two component
TL liquid to the single component TL liquid consisting of the bound state,
which is completely different from the mechanism for $\alpha<0.7$
based on the shape change of the dispersion curve of the spinon
excitation carrying $S^z=1/2$.

To summarize, we have presented the magnetic phase diagram of the
$S=1/2$ zigzag spin chain including the strongly frustrated region,
based on the 192-site results.  Although we can not eliminate the
possibility of  a ``fine structure'' appearing in the magnetization
curve beyond the resolution of the present calculation, we believe
that the phase boundaries of the cusps and plateau are correctly
obtained.  The resulting phase diagram in Fig.1 exhibits a variety of
interesting physics: The 1/3 plateau accompanying the spontaneous
breaking of the translational symmetry, the characteristic move of the
cusp singularities around the plateau and the even-odd behavior of the
magnetization curve.  In particular we have clarified two types of
cusp mechanisms: the cusp based on the shape change of the dispersion
curve and the cusp attributed to the formation of the bound state
related to the even-odd behavior of the magnetization curve.  In the
background of these behaviors, we have focused on the role of the
frustration in terms of the competition between the single and double chain
properties.  The magnetic field triggers the switching of these two
natures which induces the cusp transitions.  In addition, the 1/3 plateau
and the cusps remarkably merge with each other at $\alpha \simeq 0.7$
and $0.82$.  However, the details of the mechanism for this remain an
interesting future problem.

Although the zigzag chain is a quite simple model, we consider that it
captures the intrinsic aspects common to a class of frustrated spin
chains having the zigzag type structure. For the exploration of such models,
we believe that our result is instructive and useful.



We would like to thank T. Hikihara for valuable comments.  K.O. also
thanks A. Koga and A. Kawaguchi for fruitful discussions.  This work is
partially supported by a Grant-in-Aid for Scientific Research on
Priority Areas (B) from the Ministry of Education, Culture, Sports,
Science and Technology of Japan.


\begin{thebibliography}{99}
\bibitem{eo} Thus the true magnetization curve passes almost  through the middle of the oscillating steps.
\bibitem{cg}We have investigated the classical chain with the conjugate
gradient method. The plateau mechanism for the classical model is
essentially the same as the one in S. Miyasihta: J.Phys. Soc. Jpn. {\bf
55} (1986) 3605.
\bibitem{qp}This argument of the quasi-particle of $S^z=1$ is
applicable to a large but finite size system.
\bibitem{size} The even-odd behavior branch of the magnetization curve is expected to be described by the TL liquid made from the bound state particles. Thus we can naturally assume the $1/N$ dependence for $\Delta_N(m)$.
\end{thebibliography}
\end{document}